\documentclass{emulateapj}

\newcommand{\uv}{\mbox{$u$-$v$}}

\newcommand{\uksq}{{\,\mu{\rm K}^2}}

\shorttitle{SZA Measurement of Arcminute CMB Anisotropy}
\shortauthors{Sharp et al.}

\begin{document}
\title{A Measurement of Arcminute Anisotropy in the Cosmic Microwave Background with the Sunyaev-Zel'dovich Array}

\author{
Matthew~K.~Sharp,\altaffilmark{1,2} 
Daniel~P.~Marrone,\altaffilmark{1,3,4}
John~E.~Carlstrom,\altaffilmark{1,2,3,5}
Thomas~Culverhouse,\altaffilmark{1,3} 
Christopher~Greer,\altaffilmark{1,3} 
David~Hawkins,\altaffilmark{6}
Ryan~Hennessy,\altaffilmark{1,3} 
Marshall~Joy,\altaffilmark{7}
James~W.~Lamb,\altaffilmark{6}
Erik~M.~Leitch,\altaffilmark{1,3} 
Michael~Loh,\altaffilmark{1,2}
Amber~Miller,\altaffilmark{8,9}
Tony~Mroczkowski,\altaffilmark{8,10}
Stephen~Muchovej,\altaffilmark{6,10}
Clem~Pryke,\altaffilmark{1,3,5}
and David~Woody\altaffilmark{6}}

\email{sharp@oddjob.uchicago.edu}

\altaffiltext{1}{Kavli Institute for Cosmological Physics, University
of Chicago, Chicago, IL 60637, USA}
\altaffiltext{2}{Department of Physics, University of Chicago, Chicago, IL 60637, USA}
\altaffiltext{3}{Department of Astronomy and Astrophysics, University of Chicago, Chicago, IL 60637, USA}
\altaffiltext{4}{Jansky Fellow, National Radio Astronomy Observatory}
\altaffiltext{5}{Enrico Fermi Institute, University of Chicago, Chicago, IL 60637, USA}
\altaffiltext{6}{Owens Valley Radio Observatory, California Institute of Technology, Big Pine, CA 93513, USA} 
\altaffiltext{7}{Department of Space Science, VP62, NASA Marshall Space Flight Center, Huntsville, AL 35812, USA}
\altaffiltext{8}{Columbia Astrophysics Laboratory, Columbia University, New York, NY 10027, USA}
\altaffiltext{9}{Department of Physics, Columbia University, New York, NY 10027, USA}
\altaffiltext{10}{Department of Astronomy, Columbia University, New York, NY 10027, USA}

\slugcomment{The Astrophysical Journal, 713, 82-89 (2010)}

\begin{abstract}

We present  30~GHz measurements of the angular power spectrum of the cosmic microwave background (CMB) obtained with the Sunyaev-Zel'dovich Array. The measurements are sensitive to arcminute angular scales, where secondary anisotropy from the Sunyaev-Zel'dovich effect (SZE) is expected to dominate. For a broad bin centered at multipole 4066 we find $67^{+77}_{-50}~\uksq$, of which $26\pm 5~\uksq$ is the expected contribution from primary CMB anisotropy and $80\pm54~\uksq$ is the expected contribution from undetected radio sources. 
These results imply an upper limit of 155$\uksq$ (95\% CL) on the secondary contribution to the anisotropy in our maps.  This level of SZE anisotropy power is consistent with expectations based on recent determinations of the normalization of the matter power spectrum, i.e., $\sigma_8\sim 0.8$.

\end{abstract}


\section{Introduction}
\label{sec:intro}
Density perturbations at the epoch of recombination are imprinted onto the cosmic microwave background (CMB), leaving temperature anisotropy that has now been well-studied on a wide range of angular scales. On scales of several arcminutes and smaller, corresponding to multipole moments of $\ell\geq3000$, the level of CMB anisotropy power from primordial fluctuations is strongly suppressed by photon diffusion, and secondary sources of power, including the Sunyaev-Zel'dovich effect (SZE), are expected to play a significant role \citep[e.g.,][]{hudodel}. The SZE results from the inverse Compton scattering of CMB photons by the hot electron gas within clusters of galaxies \citep{Sunyaev1972}.   This interaction leaves a small spectral distortion in the CMB which produces anisotropy power on scales of $\ell\sim2000-10,000$, detectable as a decrement in the CMB intensity at 30~GHz.  The amplitude of this power is extremely sensitive to the history of structure formation and, specifically, to the value of $\sigma_8$, the normalization of the matter power spectrum \citep[e.g.,][]{komatsu2002}. Evidence for anisotropy above that expected from the primary CMB on these small angular scales has been detected at 30~GHz by the Cosmic Background Imager (CBI) and Berkeley Illinois Maryland Association (BIMA) experiments  \citep{Readhead2004, Dawson2006}. Recent constraints on excess power at these scales from the Arcminute Cosmology Bolometric Array Receiver (ACBAR) at 150~GHz indicate that the reported excess power is inconsistent with thermal CMB fluctuations but is consistent with SZE fluctuations \citep{ACBAR2008}.  Taken together, these measurements indicate a level of SZE anisotropy power consistent with a value of $\sigma_8$ somewhat greater than those preferred by other contemporary measurements of the parameter \citep{Voevodkin2004,WMAP5}.
In this paper, we describe a new, high-sensitivity measurement of  power in the CMB on scales ranging from  $\ell\sim3000 to 6000$ made at 30~GHz with the Sunyaev-Zel'dovich Array (SZA). 

\section{Observations}
\label{sec:obs}
\subsection{SZA Data}
\label{sec:szadata}
The CMB anisotropy data presented here were obtained with the SZA, an eight-element interferometer located at Caltech's Owens Valley
Radio Observatory near Bishop, California. The SZA antennas are equipped with
sensitive, wide-bandwidth
receivers operating at 30~GHz and 90~GHz. For these observations we used the 30~GHz receivers,
tuned to detect sky frequencies of 27$-$35~GHz. The
receivers are based on low-noise, cryogenic high electron mobility
transistor amplifiers \citep{HEMT}, with
characteristic receiver temperatures $T_{rx} \sim11-20~{\rm K}$. 
Including atmospheric and other noise contributions, the typical system
temperatures were 35$-$45~K.  The
3.5 m SZA antennas have a primary beam that is well described by a
Gaussian with a FWHM of 11\arcmin\ at the center of the 30~GHz band. Cross-correlations of the signals
from pairs of the eight antennas ({\it visibilities}) are formed in a digital
correlator, which processes the 8~GHz IF bandwidth in 16 bands of 500~MHz,
each of which is further subdivided into 17 channels of 31~MHz, allowing
rejection of narrow-band interference.

\begin{figure}
\plotone{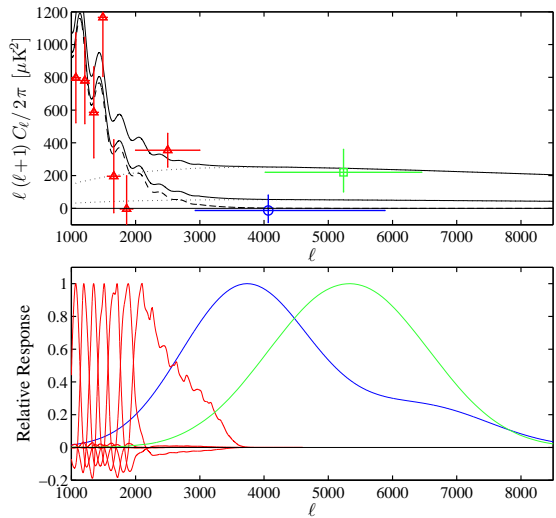}
\caption{Upper panel: CMB anisotropy power measured at 30~GHz by the SZA (blue circle, this paper), CBI (red triangles, \citep{Readhead2004}), and BIMA (green square, \citep{Dawson2006}) experiments.  Also shown are the WMAP5 primary power spectrum (dashed) and a model 30~GHz SZE spectrum (dots) for $\sigma_8$=1.0 (upper) and 0.8 (lower).  The sums of the primary and secondary power spectra are shown as solid lines. All data are corrected for contributions from undetected radio sources and error bars assume Gaussian sample variance. Lower panel: for the same experiments, we show the window functions of the measurements, normalized to their peak values.}
\label{fig:win_ps}
\end{figure}

For the anisotropy measurements reported here, the SZA antennas were arranged in a hybrid configuration to provide simultaneous sensitivity to
the arcminute-scale structure of the SZE signal from galaxy
clusters and the finer-scale contamination from radio sources \citep[details of the configuration can be found in][]{muchovej2007}.  Six of
the eight antennas were packed closely together (spacings of
4.5$-$11.5~m), yielding 15 baselines with typical projected lengths of
400$-$1400$\lambda$, corresponding to arcminute angular scales; the window function for our anisotropy measurement is
determined by the baselines formed from combinations of these six antennas.  By
convolving the distribution of projected baseline lengths in the
unflagged data with the autocorrelation of the antenna illumination
pattern, we obtain the $\ell$-space window function shown in
Figure~\ref{fig:win_ps}. This is the filter through which we observe the power spectrum; multiplying this function by $ \ell (\ell+1)C_\ell/2\pi$ and integrating over $\ell$ yields our measurement, where $C_\ell$ is the CMB angular power spectrum.  The sensitivity-weighted mean value of the window is $\ell=4066$,
with 68\% of the area encompassed by the interval $\ell=2929-5883$. In
addition to the central six antennas, two antennas were positioned
$\sim$50~m from the array center. The 13 baselines involving these antennas
are sensitive to $\sim 20$\arcsec\ scales, corresponding to multipoles of $\ell \sim 30,000$, and can be used to identify radio
sources that contaminate the short-baseline data.

Between 2005 November and 2007 June, we devoted 1340 hr of
observations to 44 SZA deep fields, comprising 2 deg$^2$. These fields were selected to pass nearly overhead at the
$\sim37^\circ$ latitude of the telescopes, and arranged in right ascension to
accommodate other large SZA observing programs. Within these
constraints, the fields were placed at locations with low Galactic
emission in the $IRAS$ 100~$\mu$m survey. No consideration was given to the
presence of radio sources
to avoid any bias that might be introduced by the correlation
between radio sources and galaxy clusters. One of the 44 fields was
found to contain a 700~mJy radio source ($>4000\sigma$); unable to exclude this source from our data with sufficient precision, we are forced
to omit this field from our analysis. Using simulated SZE sky maps we
have verified that even in the pessimistic case in which this
single field contains the most power of any of our fields, the bias
introduced by excluding it is 10\%. 

Observations were designed to allow subtraction of
ground contamination. Fields were observed at constant declination in groups of four, spaced by
four minutes of right ascension. We observed
the fields in 16 minute blocks between observations of a nearby phase
calibrator. Each field, starting with the westernmost field of the
group, was observed for just under 4 minutes at a time, so that
after including time lost to slews and calibration we tracked all four
fields through precisely the same azimuth/elevation path. This ensures
that each sees the same contribution from the ground, which can be removed by subtracting the mean of the four fields (in practice we implement this subtraction by means of a constraint matrix, described in Section ~\ref{sec:lh}). At an integration
time of 20 s, we obtain 10 integrations per field in each
cycle. The total integration time per field was $\sim25$ hr. Of
this, approximately 20\% of the data were discarded due to hardware problems, antenna shadowing, excessive noise, or jumps in
calibrator phase or system temperature which occurred on a time-scale
faster than our calibrations. An additional $\sim$5\% of the data were
flagged for showing unexpected correlations among baselines and bands
or statistically unlikely noise behavior on a single antenna.  These
cuts significantly improved the results of the jackknife tests
discussed in Section ~\ref{sec:jk}; the data quality cuts were refined using only the jackknifed data sets to avoid biasing ourselves against signal. The first member of each group of four
SZA fields is listed in Table \ref{tab:fieldTab}, along with its
position, the total, unflagged, on-source integration time, and the
achieved rms noise level for the field, both with and without data from the long baselines.

\begin{deluxetable*}{lccccc}
\tabletypesize{\scriptsize}
\tablecolumns{5}
\tablecaption{SZA Anisotropy Fields}
\tablehead{
\colhead{Group of } & \colhead{First Field RA}& \colhead{Decl.} & \colhead{Integration Time} & \colhead{rms Noise (Short Baselines)} &\colhead{rms Noise (All Baselines)} \\
\colhead{Four Fields} & \colhead{ }& \colhead{} & \colhead{(hr)} & \colhead{(mJy)} &\colhead{(mJy)}}
\startdata
cmb07  & 02:07:37.00 & +34:00:00.0 & 22.9 & 0.18 & 0.12\\
cmbCC1 & 02:11:31.30 & +33:27:43.0 & 21.9 & 0.19 & 0.14\\
cmbA1  & 02:12:00.00 & +33:00:00.0 & 19.7 & 0.21 & 0.14\\
cmbI1  & 02:12:00.00 & +32:37:08.2 & 22.8 & 0.21 & 0.14\\
cmbR1  & 02:12:15.60 & +32:11:24.8 & 32.3 & 0.18 & 0.12\\
cmbY1  & 02:12:00.00 & +31:51:24.4 & 23.0 & 0.18 & 0.12\\
cmbDD1 & 14:18:40.10 & +35:01:42.0 & 18.0 & 0.26 & 0.17\\
cmbEE1 & 14:18:39.24 & +35:31:52.3 & 20.9 & 0.19 & 0.14\\
cmbXXb & 21:28:50.60 & +24:59:35.0 & 18.2 & 0.21 & 0.15\\
cmbAA1 & 21:24:38.70 & +25:29:37.0 & 19.6 & 0.22 & 0.16\\
cmbBB1 & 21:24:38.10 & +25:59:24.0 & 18.8 & 0.21 & 0.15
\enddata
\label{tab:fieldTab}
\end{deluxetable*}

The absolute calibration of the flux density scale was derived from
bimonthly observations of Mars. We use the \citet{rudy1987} model to
predict the brightness temperature of Mars as a function of frequency
and time. Since the planet is partially resolved by our longest
baselines, a strong, unresolved source is used to transfer the
calibration from the short baselines. The absolute calibration was
cross-checked by comparing SZA observations of Jupiter to those of
\textit{Wilkinson Microwave Anisotropy Probe} ({\it WMAP}) and CBI \citep{wmapJupiter,Readhead2004}.  Based on these
measurements, we conservatively estimate that our absolute flux scale is accurate
to better than 10\% (20\% in power).

\vspace{3em}
\subsection{Very Large Array Data}
Away from the Galactic plane, the most significant contributor of
non-CMB power on arcminute angular scales at 30~GHz is compact radio
sources.  Since the power from these sources is constant as a function of $\ell$, 
their contribution to $\ell (\ell+1)C_\ell/2\pi$ can be quite large at small angular scales.  While the
brightest sources at 30~GHz can be detected by the long baselines of
the SZA, sources near our noise level (and hence undetectable in our
data) can still contribute substantial power to the measurement. We
therefore supplement our data with a higher-sensitivity search for
radio sources in our fields using the NRAO\footnote{The National Radio Astronomy Observatory is a facility of the National Science Foundation operated under cooperative agreement by Associated Universities, Inc.} Very Large Array (VLA).

The large disparity between the VLA (25~m) and SZA (3.5~m) antenna
sizes makes it impractical to survey the SZA fields at a frequency
near 30~GHz. Instead, we elected to conduct the VLA survey at 8~GHz, a
compromise between survey speed, which decreases toward higher
frequency as the VLA primary beam shrinks, and spectral extrapolation
of the detected sources. In 2006 November, we collected 36 hr of
data with the VLA in its C configuration, targeting the 24 fields that
the SZA had observed at that point.  In 2008 August, an additional set of 
observations covering all but two of
the remaining fields were obtained
in the more compact D configuration, which better matches the
angular resolution of the SZA long baselines.  Because the VLA primary beam at
8~GHz is only 4.9\arcmin\ across, it was necessary to make mosaic
observations with 19 pointings to cover the area of the SZA's larger
primary beam. 

 We reached a noise level of $\sim30~\mu$Jy at
the center of this mosaic pattern, less than 1/6 of the noise level in
our SZA data.  The integration time is tapered at pointings away from
the center of the mosaic, such that the achieved sensitivity in the
8~GHz maps has a profile that matches that of the 30~GHz data, 
set by the SZA primary beam pattern.

The VLA data were calibrated and mosaicked in AIPS. The map
sensitivity $\sigma$ was determined from the mosaic images using the task RMSD.
Sources were
then extracted from the maps using the task SAD, with
a limiting significance of 5$\sigma$, yielding more than 180 sources.

Due to weather and scheduling limitations, two of the SZA fields
were not observed with the VLA at 8~GHz.  We omit the entire group of four fields that includes this pair from the analysis that follows, in addition to the single field that contained a bright radio source.  This leaves us with 39 of the 44 observed fields to include in our analysis.

\section{Analysis}
\label{sec:lh}
\subsection{Likelihood Analysis}
As an interferometer, the SZA directly measures the Fourier components of the sky brightness.  Since it is precisely the variance of these components that the power spectrum describes, interferometers are well suited for measuring the power spectrum, without the intervening stage of map making \citep[e.g.,][]{White1999}.

We have used a maximum likelihood method to extract measurements of the power spectrum from our data.  Following \citet{Bond1998}, we construct a vector $\Delta$ of $N$ visibilities and a likelihood estimator that is a function of $\Delta$ and the CMB power $\kappa$, assumed to be constant across the range of angular scales probed by the SZA.  We expect our data, and thus the likelihood function, to be Gaussian, so that the likelihood is given by:
\begin{equation}
\label{eqn:LH}
\mathcal{L}(C)=\frac{1}{(2\pi)^{N/2}|C|^{1/2}}\exp\left(-\frac{1}{2}\Delta^T C^{-1} \Delta\right)
\end{equation}
where N is the length of the data vector $\Delta$, and the covariance $C$ of the data is written as the sum of a contribution from the CMB, the diagonal instrument noise, and a set of constraints (see below):  
\begin{equation}
\label{eqn:covar}
C=\kappa\,C_\mathrm{CMB}+C_\mathrm{noise}+C+\mathrm{constraints}.
\end{equation}

If our visibilities were independent of one another, the contribution to the covariance matrix from the CMB signal, $\kappa\,C_\mathrm{CMB}$, would be the identity matrix times the level of CMB power, $\kappa$.  We assume a flat band power, i.e., constant $\ell (\ell+1)C_\ell/2\pi$.   However, in practice, visibilities can be correlated, depending on the separation of the Fourier-space (\uv) coordinates they sample.  A visibility corresponding to a given \uv\ coordinate is in fact an integral over a small patch of the Fourier plane, weighted by the antenna's aperture autocorrelation function; the overlap integral of this function centered on two neighboring \uv\ coordinates yields the correlation between adjacent measurements.  We calculate these overlap integrals for each of our visibilities and insert the resulting matrix into the total covariance matrix as indicated in Equation (\ref{eqn:covar}).  Visibilities that are more than 95\% correlated are averaged together.

The instrumental contribution to the noise, $C_\mathrm{noise}$, dominates the weak signal from the CMB anisotropy and must therefore be accurately determined.  We estimate the noise variance from the scatter of visibilities taken on 20 s timescales within a 4 minute scan.  From the 10 visibilities within each scan, we subtract the mean of the real and imaginary parts from each sample and then compute a single sample variance from the 20 mean-subtracted numbers.  All averages within the analysis are uniformly weighted, as weighting by variances derived from just 20 samples introduces substantial biases.  The resulting noise estimates are independent of one another, so $C_\mathrm{noise}$ in Equation (\ref{eqn:covar}) is diagonal.  We have verified through simulation that variances measured in this way produce an unbiased estimate of the noise, and, hence of the CMB power level.

\subsection{Constraining Radio Sources}
We expect the sky signal to be dominated by radio sources that are point-like at the SZA's arcminute resolution, principally active galactic nuclei emitting synchrotron radiation at radio frequencies.    Through the use of constraint matrices \citep{Bond1998, Halverson2002}, we can eliminate the modes of the data that are corrupted by these sources, provided that we know their positions.  The constraint matrix is computed by writing a template vector of visibilities that describes the contribution of the contaminant in question to each \uv\ component; in the case of a point source at position $(l,m)$ with respect to the interferometer's phase center, the contribution $V_{pt}^j$ to the visibility measured with component $(u_j,v_j)$ is given by:
\begin{equation}
V_{pt}^j=Ae^{2\pi i (u_j l + v_j m)},
\end{equation}
where $A$ is an unknown amplitude. By forming the outer product of this vector with itself, $C^{ij}_{pt} = V^i_{pt} V^j_{pt}$, we find the covariance created among our $N$ visibilities by a source at this position; the amplitude of this covariance remains unknown, but its shape is fixed by the (known) source location.  Each constraint matrix is then multiplied by a large prefactor (as large as possible without causing the matrix to become poorly conditioned for inversion) and added to the total covariance matrix in our likelihood function, effectively setting to zero the weight of the mode corresponding to the source.  The unknown amplitude is therefore unimportant; the operation is equivalent to marginalizing over the source's unknown flux.  We form such a matrix for each source that we wish to eliminate from our data; because each matrix is created from a single vector, each has rank one and eliminates 1 degree of freedom.   Radio sources in our data may have a wide range of spectral indices and cannot in general be excluded by a single constraint. Instead, we project 3 modes of different spectral index for each source; we therefore lose three degrees of freedom for each projected source.  We have verified in simulation that a linear combination of these components can effectively remove the contribution of a source of any reasonable spectral index. 

We remove a component common to the fields within a group of four (ground or antenna crosstalk) using a similar constraint.  This constraint identifies each visibility among a group of four with its three counterparts in different fields, all of which are measured at the same antenna position.  Projecting this constraint is equivalent to subtracting the mean of the four visibilities from each of them, and therefore reduces the sensitivity by 25\%.

The radio source positions are derived from several data sets. The brightest sources at 30~GHz are identifiable directly in the SZA data. For purposes of source detection, we combine the long and short-baseline data (typical noise of $\sim150~\mu$Jy), providing 30\% better sensitivity to compact sources than would be obtained using only the short baselines from which our anisotropy measurement is determined.   Above a
significance of $5\sigma$ in the 30~GHz data, 42 of 44 detected sources have counterparts in the 8~GHz VLA survey; the remaining two sources were found to be extended even at the 20\arcsec\ resolution of the long SZA baselines and are likely heavily resolved by the VLA in its C 
 configuration (3\arcsec\ resolution). Both of these sources have 1.4~GHz counterparts in the NRAO VLA Sky Survey (NVSS; \citealt{NVSS}), which has angular resolution intermediate to that of the long and short SZA baselines, giving us confidence that they are real, extended sources rather than noise peaks in our 30 GHz maps. 

As sources below our 30~GHz detection threshold can still contribute significant anisotropy power, we constrain 180 sources detected at 5$\sigma$ in the 8~GHz survey, as well as the two sources detected at 30~GHz that are too resolved to be seen with the VLA.  The effect of using different detection thresholds in the two data sets can be seen in Table~\ref{tab:pntsrc}.  Clearly, radio sources contribute the bulk of the detected anisotropy; even sources just above the 5$\sigma$ threshold at either frequency contribute substantial power.  We also note that the two sources without 8~GHz counterparts contribute a sizable fraction of our final power; the inclusion of these two sources is therefore important to our final result.

\begin{deluxetable}{lccc}
\tablecolumns{4}
\tablewidth{0pt}
\tablecaption{Power Measured Using Various Radio Source Catalogs\label{tab:pntsrc}}
\tablehead{30~GHz SZA & VLA 8~GHz & Power & Number of Sources \\
Threshold & Threshold & ($\mu$K$^2$) & Constrained}
\startdata
6$\sigma$ & \nodata &$190^{+81}_{-62}$&33 \\
5$\sigma$ & \nodata &$115^{+79}_{-52}$&44 \\

\hline
\nodata & 20$\sigma$ &$315^{+92}_{-64}$ &  81 \\
\nodata & 15$\sigma$ &$235^{+84}_{-65}$ &  98 \\
\nodata & 8$\sigma$  &$115^{+80}_{-54}$& 141 \\
\nodata & 5$\sigma$  &$103^{+75}_{-56}$& 180 \\
\hline
5$\sigma$ & 5$\sigma$ &  $\phantom{0}67^{+77}_{-50}$  & 182 
\enddata
\tablecomments{Measurements include 39 fields.}
\end{deluxetable}

 For each field, we assemble the noise and constraint components of the covariance matrix and vary the level of CMB power, evaluating the likelihood of the data according to Equation (\ref{eqn:LH}) over a range of values for the power $\kappa$.  Treating our fields as independent samples, we take the product of all of their likelihood curves to form a global likelihood for the experiment; we report the maximum of this curve as the most likely power, and we use the points that correspond to 16\% and 84\% of the cumulative likelihood to define the 68\% confidence interval.  The likelihood is allowed to extend below zero power.

We tested this analysis pipeline extensively with simulated data, including tests in which the data generation and power spectrum analysis were performed by different parties, using independent software packages, and  found consistent results in all cases.
 
\subsection{Jackknife Tests}
\label{sec:jk}
Processing the data as described above, we can compute the relative likelihood of different levels of CMB power, given our data. We now subject the data to three jackknife tests to rule out contaminating power from non-astronomical sources or from inaccuracies in our noise variance estimation. In the jackknife tests, the data are split into halves, each of which measures very nearly the same set of spatial Fourier components.  The data are taken in such a way that three of these splits are possible: (1) a frequency jackknife, wherein we separate alternating 500 MHz frequency bands, as neighboring bands sample very nearly the same \uv\ points, (2) a jackknife in which the data are separated into halves by time, and (3) a jackknife in which 
the data are separated into even and odd days. The latter two jackknife tests take advantage of the fact that the SZA observed the same fields in the same way for many days.

After dividing the data into halves, we difference the matching pairs of visibilities. Emission from astronomical sources, which is common to both halves, is removed, while any contamination that varies with time or frequency will remain in the differenced (jackknifed) data. Note that the requirement that there be matching, unflagged visibilities in both halves for each test results in some additional loss of data and degradation of the measurement sensitivity, particularly in the two time-jackknife tests. We compute variances from the differenced visibilities, and compute the power in each jackknifed data set just as for our unjackknifed data, with the expectation that the measured power will be consistent with zero.
The jackknife tests also verify our noise model; if the noise in our data has been misestimated, we expect that the jackknifed data will be poorly described by our computed covariance matrices.  

The jackknife tests provide evidence that we detect a signal correlated with the orientation of our antennas relative to the ground. This signal may originate from antenna crosstalk or ground emission. As discussed in \S~\ref{sec:szadata}, the observations were designed to allow removal of such ``ground contamination" by subtracting the average of each group of four observations via an additional constraint. We find that without removing this component (but including strict cuts on data quality discussed in \S~\ref{sec:szadata}) the frequency jackknife test fails badly, with a probability to exceed the jackknifed $\chi^2$ of $10^{-5}$. After implementing the ground subtraction, the three jackknife tests all pass, as seen in Table \ref{tab:jk}, with probabilities to exceed the $\chi^2$ greater than 0.1.  We therefore employ ground subtraction to determine anisotropy power in all the analyses presented in this paper.  Our analysis procedure is blind to the final result, as we have refined our data cuts and analysis treatment based only on the jackknife tests.

\begin{deluxetable}{lccc}
\tablecolumns{4}
\tablewidth{0pt}
\tablecaption{Anisotropy Measurements for Raw and Jackknifed Data\label{tab:jk}}
\tablehead{Jackknife Test & Power ($\mu$K$^2$) & PTE}
\startdata
Frequency          &  $\phantom{-}46^{+66\phantom{0}}_{-44}$ &  0.27\\
First--Second Half &  $\phantom{-0}4^{+126}_{-88}$ &  0.56\\
Even--Odd Days     & $\phantom{-}13^{+92\phantom{0}}_{-66}$ &  0.59\\
\hline
Unjackknifed Data  & $\phantom{-}67^{+77\phantom{0}}_{-50}$& 0.10
\enddata
\tablecomments{Results from 39 fields with the removal of sources detected at 5$\sigma$ at 8 or 30~GHz and ground subtraction. }
\end{deluxetable}

\section{Anisotropy Contributions}
\label{sec:res}
Repeating the analysis on the unjackknifed data, including ground-subtraction and constraining radio sources detected at $>5\sigma$ in both the 8 GHz VLA data and the 30 GHz SZA data, leads to a residual power of $67^{+77}_{-50}\uksq$. To set constraints on secondary CMB anisotropy, we must estimate the residual power contributions to this measurement from other astronomical sources. In this section, we first consider contributions from primary CMB anisotropy, undetected radio sources, and diffuse Galactic emission, and we end with our constraint on the level of secondary anisotropy.  We note that contributions from undetected radio sources and diffuse Galactic emission 
are, in principle, distinguishable from the CMB by their spectral signature across the 25\% fractional bandwidth of the SZA data. However, it is not possible to distinguish a preferred spectral index given the low signal-to-noise ratio of the data.

\subsection{Primary CMB}
At a multipole of $\ell=4000$, the WMAP5 cosmology \citep{WMAP5} predicts no significant primary CMB anisotropy (Figure~\ref{fig:win_ps}). However, our window function has limited sensitivity to the larger scales at which the primary CMB signal is strong. We estimate this contribution and its variance with simulated observations. We generate CMB skies according to WMAP5 cosmological parameters using CMBFAST \citep{CMBFAST}. We use 35 sets of 39 fields with identical noise realizations but different CMB realizations and sample them according to the SZA \uv\ coverage and noise. From these simulations, we determine that the primary CMB contributes a mean power of $26 \pm 5\uksq$ to our measurement. Integrating the CMB primary power spectrum multiplied by our window function yields a very similar estimate.

\subsection{Undetected Radio Sources}
\label{sec:radsrcs}
To assess the level of residual fluctuations from sources that are undetected at 8~GHz and 30~GHz at our 5$\sigma$ threshold (see \S~\ref{sec:lh}), we repeat our analysis using simulated data that include a 30~GHz selected source population extending below our detection thresholds.  This population's source density as a function of flux is derived from the SZA blind cluster survey \citep{muchovej2009}, which provides source counts down to 1~mJy; we extrapolate to lower flux levels according to a power law. This survey is the deepest available at this frequency.  To assign 8~GHz fluxes to our simulated population of sources, we use the spectral index distribution of \citet{muchovej2009} measured for 200 sources between 5 and 30~GHz. We simulate SZA observations of fields containing these source populations, constraining those that appear above our detection thresholds at either 8 or 30~GHz. We find that undetected sources in measurements of 39 fields contribute $80\pm54~\uksq$ to the detected power.

\subsection{Galactic Synchrotron Radiation and Free-Free Emission}
At lower frequencies, emission from the Galaxy via synchrotron and free-free radiation contribute significantly to the sky brightness, as can be seen in the {\it WMAP} 22~GHz all-sky map \citep{Gold2008}. We lack sensitive, high-resolution measurements of these components and are forced to estimate their contribution from lower-resolution maps and forecasts derived from them. \citet{Tegmark2000} used degree-resolution templates to predict that this emission should contribute less than $1~\uksq$ at $\ell\sim4000$. Renormalizing their model using an analysis of foregrounds measured with WMAP \citep{Tegmark2004} increases this estimate to $2~\uksq$ at 30~GHz.

\subsection{Galactic Spinning Dust}
An additional source of potential foreground contamination in our power spectrum measurement comes from Galactic dust.  While the predictions of \citet{FDS} at 30~GHz imply $<1\uksq$ contribution for thermal dust emission, there has been evidence that dipole radiation from small, spinning dust grains may dominate the total dust emission at 30~GHz.  Such emission is expected to peak at frequencies of tens of GHz and to be tightly correlated with Galactic dust \citep{Draine1998}.  Dust-correlated emission within this frequency range has been observed on various angular scales \citep{Leitch1997,Kogut1996,tenerife2002,finkbeiner2002,Finkbeiner2004,Watson2005,Dickinson2007}.  During the SZA blind cluster survey, increased map noise was indeed observed in fields with more obvious {\it IRAS} structure.

To estimate the contribution of spinning dust to the SZA measurement, we calculate the power spectrum of the {\it IRAS} $100$ $\mu m$ maps of our fields, fit it to $\log_{10}\mathrm{Power}=A\ell +B$, and extrapolate it to the smaller angular scales that we are sensitive to, but to which {\it IRAS} is not.  We then take the highest measured emissivities of dust-associated emission at 30~GHz, as tabulated by \citet{Dickinson2007}, and scale this power spectrum to match the observations.  Multiplying by the window function shown in Figure \ref{fig:win_ps} and integrating over $\ell$, we get a contribution for each of the observed emissivities lying in the range from $<1~\uksq$ to $16~\uksq$.  Note that the highest emissivity \citep{Leitch1997}, derived from a very localized portion of the Galactic plane, is substantially higher than other measurements, and so the corresponding $16~\uksq$ is likely a substantial overestimate of the contribution of this foreground to our measurement, made in a different part of the sky with less Galactic emission.

\subsection{Secondary CMB Anisotropy}

We now estimate the contribution from secondary CMB anisotropy.  We account for the 
expected contributions from primary CMB anisotropy and undetected radio sources. We do not consider the contributions from sources of diffuse Galactic emission, since even the conservative estimates discussed above are subdominant to the contributions from primary CMB and undetected radio sources.

Subtracting the residual power for undetected radio sources and including our 20\% calibration uncertainty, we are left with $-13^{+94}_{-74}\uksq$ for the combined primary and secondary anisotropy result.  The  constraint on the level of secondary CMB anisotropy after subtracting the expected contribution from primary CMB anisotropy is $-39^{+94}_{-74}$. Applying a prior of positive power
and integrating the likelihood curve results in an upper limit to the level of secondary CMB power in our maps of $155 \uksq$ at 95\% CL ($72 \uksq$ at 68\%). 

The estimated anisotropy contributions and the resulting constraint on the level of secondary anisotropy are tabulated in Table \ref{tab:foregrounds}. In Figure~\ref{fig:win_ps}, we plot the SZA anisotropy measurement corrected only for residual power from radio sources but including the contribution from primary CMB, along with previous measurements of small-scale CMB power at 30 GHz from CBI and BIMA \citep[][respectively]{Readhead2004, Dawson2006}.

\begin{deluxetable}{lc}
\tablecolumns{2}
\tablewidth{0pt}
\setlength{\tabcolsep}{5mm}

\tablecaption{Anisotropy Contributions}
\tablehead{
\colhead{Source}&\colhead{Power Contribution ($\mu$K$^2$)}
}

\startdata
Primary CMB & $26\pm5$\\
Undetected Radio Sources & $80\pm54$\\
Galactic Synchrotron and Free-Free & 2 \\
Galactic Spinning Dust & $<16$\\
\hline
Secondary CMB (SZE) & $-39^{+94}_{-74}$\\
\quad and upper limit at 95\% CL & 155\\
\enddata
\label{tab:foregrounds}
\end{deluxetable}

\section{Discussion}
\subsection{Constraints on $\sigma_8$ from cluster simulations}
To understand the implications of our measurement for the value of $\sigma_8$, we compare it to mock observations of simulated SZE maps.  Several  groups have carried out large-scale cosmological structure simulations and converted three-dimensional, simulated universes into two-dimensional projections of the Compton $y$-parameter, the frequency-independent measure of the magnitude of the SZE. We take these maps as inputs to mock SZA observations, and attempt to estimate the mean level of power that would be measured by the SZA, as well as the scatter in these measurements. 

We use 60 maps, each $0.5^\circ\times0.5^\circ$, from the hydrodynamical simulation of \citet{White2002}, generated with $\sigma_8=0.9$, including cooling and feedback.  \citet{Schultz03}  provide 360 maps from dark matter simulations with $\sigma_8=1.0$ and gas pasted into cluster haloes after the simulation is complete; \citet{Shaw2007} follow a similar method, producing 100 maps with $\sigma_8=0.77$.   Finally, \citet{Holder2007} use the Pinocchio algorithm of \citet{Monaco2002} to generate halo distributions and merging histories, resulting in 900 maps over a range of $\sigma_8$. 

For each set of simulated maps, we form 50 groups of 39 fields; we pick the maps from the available set randomly with replacement.  For each map, we simulate SZA observations, reproducing the \uv\ coverage and noise properties of the actual data.  We process these mock observations just as we do the real data, and produce a maximum likelihood power for each set of 39 fields.  We compute the mean and scatter of these powers from the 50 independent realizations.  

\begin{figure}
\plotone{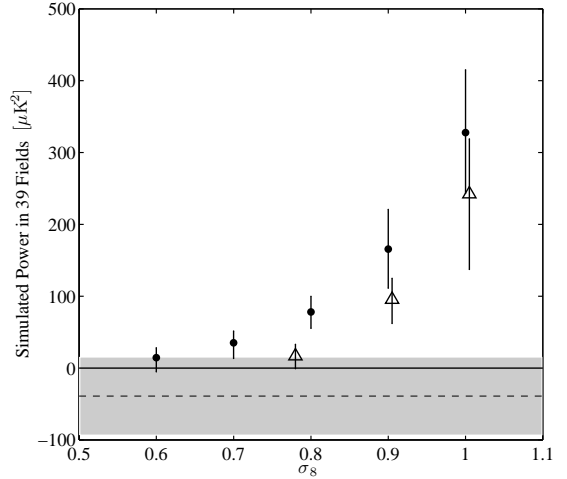}
\caption{Mean and scatter of the resultant maximum likelihood power of sets of 50 simulated realizations of the SZA measurement, including appropriate instrumental noise.  Each set uses simulated Compton $y$-maps made with a specific technique and a specified $\sigma_8$.  Four different sets of input $y$-maps were used: solid dots at five values of $\sigma_8$ (0.6$--$1.0) from \citet{Holder2007}, triangles at $\sigma_8$=0.77, 0.9, and 1.0 from \citet{Shaw2007},  \citet{White2002}, and \citet{Schultz03}, respectively. The points at $\sigma_8$ = 0.9 and 1.0 are offset slightly from those values for clarity.  The error bars reflect the scatter of the realizations, and therefore account for the noise of the measurement as well as the additional sample variance from the non-Gaussianity of the $y$-maps.  The dashed line and shaded region represent the maximum likelihood power for the secondary CMB anisotropy measured by the SZA, including only the uncertainty due to the absolute calibration and the uncertainty in the subtraction of power from primary CMB anisotropy and undetected radio sources. 
}
\label{fig:sim_power}
\end{figure}

Results for each set of simulated maps are shown in Figure \ref{fig:sim_power}.  The strong dependence of the detected power on the value of $\sigma_8$ is clear,  as are significant systematic differences among simulations.  The maximum likelihood power for the secondary CMB anisotropy measured by the SZA is shown by a shaded horizontal bar in the figure. The shaded bar indicates the 68\% confidence interval, including the uncertainty in the absolute calibration and the subtraction of power from primary CMB anisotropy and undetected radio sources. 

Figure \ref{fig:sim_power} shows that the range of $\sigma_8$ values consistent with the SZA anisotropy measurement is large, reflecting significant differences among the simulations.  The figure shows there is no tension between the SZA anisotropy measurement with $\sigma_8\sim0.8$ determined from recent measurements using other techniques, while values higher
than 0.9 are disfavored.

We note that the power spectra derived from simulated $y$-maps have greater sample variance than maps of Gaussian noise with the same rms power.  While we can measure the variance in simulation by measuring the scatter of many realizations of our experiment, in the actual data we assume Gaussianity in calculating our error bars.  We are therefore underestimating the sample variance in our measurement.  Using the simulated $y$-maps, we find that the magnitude of this underestimate has a weak dependence on the value of $\sigma_8$, but for $\sigma_8 \sim 0.8$ the actual confidence region of our measurement is likely 1.5 times broader than the range reported here, which assumes Gaussian sample variance.

\subsection{Correlations Between Radio Sources and Clusters}
There is observational evidence that radio sources are spatially correlated with clusters (e.g., \citealt{Coble2007}; \citealt{Lin2007}; and references therein). Because the short SZA baselines are sensitive to both compact radio sources and the cluster SZ signal, the projection of cluster-correlated radio sources could, in principle, remove significant secondary anisotropy along with the radio source power, biasing our measurement low. 

We examine the magnitude of this effect by simulating our measurement. We simulate observations of sets of 39 Compton $y$-maps from \citet{White2002} including appropriate instrumental noise. For these fields we generate radio source populations according to the prescription of section~\ref{sec:radsrcs} and include constraints for those that would be detected in our 8 or 30~GHz data. We introduce a pessimistic degree of correlation between clusters and radio sources by increasing the radio source density in the inner 0.5\arcmin\ of the simulated clusters according to the excess observed toward massive ($\sim10^{15} M_\odot$) clusters by \citet{Coble2007}. We find that the mean measured power in these maps is reduced by 50\%. This should be a significant overestimate of the effect in our observations as the signal we seek comes from clusters an order of magnitude less massive than those observed by \citet{Coble2007}, with proportionately fewer radio sources per halo. 

Although the short SZA baselines are unable to discriminate between radio sources and clusters, the long baseline data are wellsuited to making this distinction and could be used to recover the power lost to projection of correlated sources. Our simulations show that the inclusion of long baselines in our analysis restores an average of 25\% of the lost power. However, for any particular realization of the data the difference between the power measured with and without the long baselines may be positive or negative, with only a slight preference for increased power for our noise level and given the $110\uksq$ mean power in the simulated maps. We therefore conclude that for a single realization of our observation (the real data), a comparison of the anisotropy with and without the long baselines is not a useful predictor of the fraction of power lost to source-cluster correlation. 

\subsection{Comparison with CBI and BIMA}
Previous measurements at 30~GHz by CBI \citep{Readhead2004} and BIMA \citep{Dawson2006} have suggested power at high $\ell$ in excess of the primary CMB anisotropy, implying a value of $\sigma_8$ inconsistent with other contemporary measurements of the parameter if the excess power is attributed to SZ signal.  The CBI experiment reported 
$355^{+137}_{-122} \uksq$ for their high-$\ell$ bin spanning $2000 < \ell < 3500$, of which 
80$-$90~$\uksq$ is attributed to primary CMB anisotropy. The BIMA experiment reported $220^{+140}_{-120}  \uksq$ for $4000 < \ell < 6500$, at which there is no significant primary CMB contribution.   The SZA data presented here do not show evidence of excess power. However, the discrepancy between the SZA result and the CBI and BIMA results is not highly significant after accounting for measurement uncertainties and particularly the non-Gaussian sample variance discussed above.  Here,  we investigate whether the discrepancies may also be partially accounted for
by the different prescriptions used to account for residual radio source power.

For the CBI experiment, \citet{Readhead2004} used NVSS to remove sources with a limiting 1.4~GHz flux of 3.4~mJy from their 30~GHz anisotropy data. They estimated the residual contribution from undetected sources to be $\sim20\%$ of their highest-$\ell$ bandpower, or $\sim90\uksq$, with an uncertainty of $48\uksq$. 
Repeating our SZA analysis using the same prescription, i.e., constraining only sources detected at $> 3.4$~mJy in the NVSS survey, results in an increase of our measured power by $408^{+127}_{-105} \uksq$, raising our total residual power to $488^{+138}_{-118}$. Extrapolating this to the $\ell_\mathrm{eff}$ of the CBI measurement according to the respective window functions of the two experiments implies a residual power of
$152^{+43}_{-37}\uksq$ for the CBI result, only slightly higher than reported and consistent within the uncertainties.  The discrepancy between the measurement presented here and that of CBI is therefore unlikely to be due to the different prescriptions for estimating residual point source power.

The BIMA experiment \citep{Dawson2006} used a 4.8~GHz VLA survey, with depth comparable to our 8~GHz survey, to generate source constraints for their anisotropy measurement.  Based on the flattening of measured power versus VLA detection threshold, they estimated that radio sources contribute negligible residual power. 
We note that the flattening of the measured power with detection threshold could indicate other sources of contamination not constrained by their observational strategy, such as ground pickup. The low significance of the BIMA result, especially in light of the large sample variance resulting from the small sky area covered by the measurement, implies no significant tension with the result presented here.

\section{Conclusions}

We present results from 30~GHz measurements of the CMB angular power spectrum with the SZA, on scales where the secondary anisotropy from the SZE is expected to dominate. For a broad bin centered at multipole 4066, we find $67^{+77}_{-50}~\uksq$, of which $26\pm 5~\uksq$ is the expected contribution from primary CMB anisotropy and $80\pm54~\uksq$ is the expected contribution from undetected radio sources. The resulting constraint on secondary anisotropy is $-39^{+94}_{-74}$, implying an upper limit of $155 \uksq$ at 95\% CL ($72 \uksq$ at 68\%).  The SZA results indicate lower secondary anisotropy power than previously reported at 30~GHz by 
CBI \citep{Readhead2004} and BIMA \citep{Dawson2006}.   The discrepancies between the SZA result and the CBI and BIMA results,  however, are not highly significant after accounting for measurement uncertainties and non-Gaussian sample variance.

We show that the level of SZE anisotropy power implied by the SZA measurement is in good agreement with expectations based on simulated Compton $y$-maps made with $\sigma_8\sim 0.8$, but disfavor values of $\sigma_8 \ge 0.9$. The differences among various simulations, however, prevent a more quantitative determination of $\sigma_8$.

\bigskip
\acknowledgments

We thank John Cartwright, Ben Reddall, and Marcus Runyan for their  significant contributions 
to the construction and commissioning of the SZA instrument.  We thank the staff of the Owens Valley Radio Observatory and CARMA for their outstanding support.  We also thank Tom Crawford, Kyle Dawson, Nils Halverson, Bill Holzapfel, Ryan Keisler, Tom Plagge, Tony Readhead, and Jonathan Sievers for helpful discussions, and David S. Meier and Eric Greisen for assistance with the VLA data reduction.
We gratefully acknowledge the James S.\ McDonnell Foundation, the National Science Foundation 
and the University of Chicago for funding to construct the SZA.  
The operation of the SZA is supported by NSF Division of Astronomical Sciences through
grant AST-0604982. Partial support is provided by NSF Physics Frontier Center grant PHY-0114422 
to the Kavli Institute of Cosmological Physics at the University of Chicago, and by NSF grants 
AST-0507545 and AST-05-07161 to Columbia University.  AM acknowledges support from a Sloan Fellowship, 
SM from an NSF Astronomy and Astrophysics Fellowship, and CG, SM, and MS from NSF Graduate Research Fellowships. 

{\it Facilities:} \facility{SZA}, \facility{VLA}


\end{document}